\documentstyle[12pt,epsfig]{article} 
\input{psfig}
\newlength{\dinwidth}                       
\newlength{\dinmargin}                      
\def\lsim{\mathrel{\rlap{\lower4pt\hbox{\hskip1pt$\sim$}}
    \raise1pt\hbox{$<$}}}                
\def\gsim{\mathrel{\rlap{\lower4pt\hbox{\hskip1pt$\sim$}}
    \raise1pt\hbox{$>$}}}                
\begin{document}
\sloppy
\thispagestyle{empty}
\begin{titlepage}

\begin{flushleft}
{\tt hep-ph/9609489} \\[0.1cm]
September 1996
\end{flushleft}
\vspace{2cm}
\begin{center}
\Large
{\bf Future Precision
 Measurements \\
 of $F_2(x,Q^2)$,  $\alpha_S(Q^2)$ and $xg(x,Q^2)$ 
at HERA 
} \\
\vspace{1.2cm}
  \begin{large}
M. Botje$^a$, M. Klein$^b$, C. Pascaud$^c$ \\
  \end{large}
\normalsize
\vspace{1cm}
$^a$ NIKHEF PO Box 41882, NL-1009 DB Amsterdam, Netherlands\\
$^b$ DESY-IfH Zeuthen, Platanenallee 6, D-15738 Zeuthen, Germany \\
$^c$ Universit\'e de Paris Sud, LAL, F-91405 Orsay, France \\
\vspace{8cm}
\large
{\bf Abstract}
\normalsize
\end{center}
\vspace{0.3cm}
 The results are presented of a study of the 
accuracy one may achieve at HERA in measuring the strong coupling
constant $\alpha_{s}$ and the gluon distribution $xg(x,Q^{2})$ using
future data of the structure function $F_{2}(x,Q^{2})$ which are 
estimated to be accurate at the few \% level over the full accessible
kinematic region down to $x \simeq 10^{-5}$ and up to $Q^{2} \simeq 
50000$~GeV$^{2}$. The analysis includes simulated proton and deuteron 
data, and the effect of combining  HERA data with  fixed 
target data is discussed.
\vfill 
\noindent
\small
{\it To appear in the proceedings of the workshop ``Future Physics at 
HERA'', DESY, Hamburg, 1996. }
\normalsize
 
\end{titlepage}
%
\mbox{}
\vspace*{1cm}
\begin{center}  \begin{Large} \begin{bf}
Future Precision
 Measurements \\
 of $F_2(x,Q^2)$,  $\alpha_S(Q^2)$ and $xg(x,Q^2)$ 
at HERA \\
  \end{bf}  \end{Large}
  \vspace*{8mm}
  \begin{large}
M. Botje$^a$, M. Klein$^b$, C. Pascaud$^c$\\
  \end{large}
\vspace*{0.9cm}
$^a$ NIKHEF PO Box 41882, NL-1009 DB Amsterdam, Netherlands\\
$^b$ DESY-IfH Zeuthen, Platanenallee 6, D-15738 Zeuthen, Germany \\
$^c$ Universite de Paris Sud, LPTHE, F-91405 Orsay, France
\end{center}
\vspace*{1.5cm}
\begin{quotation}
\noindent
{\bf Abstract:} The results are presented of a study of the 
accuracy one may achieve at HERA in measuring the strong coupling
constant $\alpha_{s}$ and the gluon distribution $xg(x,Q^{2})$ using
future data of the structure function $F_{2}(x,Q^{2})$ which are 
estimated to be accurate at the few \% level over the full accessible
kinematic region down to $x \simeq 10^{-5}$ and up to $Q^{2} \simeq 
50000$~GeV$^{2}$. The analysis includes simulated proton and deuteron 
data, and the effect of combining  HERA data with  fixed 
target data is discussed.
\end{quotation}
%
\section{Introduction}
Deep inelastic scattering is the ideal place to investigate the 
quark-gluon interaction. Previous fixed target experiments have lead
to very precise tests of Quantum Chromodynamics in the kinematic
range of larger $x \geq 0.005$ and lower $Q^{2} \leq 300$~GeV$^{2}$.
The first few years of experimentation at HERA extended this range
to very low $x \simeq 0.0001$ and large $Q^{2} \simeq 3000$~GeV$^{2}$
leading to remarkable results in the investigation of deep inelastic
scattering \cite{mk,ry} including rather accurate measurements already
of the proton structure function $F_{2}(x,Q^{2})$.
In this study an attempt has been made 
to estimate the accuracy of future measurements of $F_{2}$ at HERA and
their possible impact on precision mesurements of the strong coupling
constant $\alpha_{s}(Q^{2})$ and the gluon distribution $xg(x,Q^{2})$.
The measurement of these quantities is a key task at HERA. Both can
be determined in a number of different processes as deep inelastic
jet production, charm and $J/\psi$ production and with
future measurements of the longitudinal structure function. The
measurement of $F_{2}$, however, is expected to be
the most precise  way to determine $\alpha_{s}$ and $xg$ from 
the scaling violations of $F_{2}$. Those are most prominent
at very low $x$ due to quark pair production from the gluon field and
weaker at large $x \geq 0.1 $ due to gluon bremsstrahlung. Both
processes, and their NLO corrections, will be accessible with
future high statistics data at HERA which is hoped to deliver a
final luminosity figure near to ${\cal L} \simeq 1$~fb$^{-1}$ 
during the next 8 years of operation. 

The QCD analysis of the past and present $F_{2}$ structure function 
data lead to remarkable results already, more than listed here: 
\begin{itemize}
\item{ A rather 
precise determination of $\alpha_{s}(Q^{2})$ with an experimental
error of 0.003 at $Q^{2}=M_{Z}^{2}$ was performed using the 
SLAC and the BCDMS structure function data \cite{MARCALAIN}.}
\item{ 
Both H1 \cite{h1f,h1g} and ZEUS \cite{zeusf,ZEUSQCD} have determined
the gluon distribution with an about 15\% accuracy 
at $Q^{2} = 20$~GeV$^{2}$ and $x \simeq 10^{-4}$ by using 
different sets of fixed target data \cite{SLAC,BCDMS,NMC}
combined with the HERA results}.
\item{The HERA deep inelastic structure function data have a big
impact on global analyses and the determination of parton
distributions \cite{trv}.}
\end{itemize}
The analysis presented in this paper will show that HERA will allow to
reach the 1\% level of determining $\alpha_{s}$ and $xg$.
This represents a challenge to the theoretical
understanding of deep inelastic scattering in perturbative QCD in the 
low $x$ and low $Q^{2} \sim M_{p}^{2}$ region. A precision 
measurement of the strong coupling constant will represent an 
important constraint to unified theories. As such it
represents one fundamental reason to perform an extended long term 
programme of experimentation at HERA. 

This paper is organized as follows. Section 2 presents the 
assumptions and the results of the simulation of $F_{2}$ 
structure function data. Section 3 contains the outline of the QCD 
analysis procedure and error treatment required for the analysis. 
The results of a detailed study of the $\alpha_{s}$ measurement 
accuracy are given in section 4. Similarly the determination of the 
gluon distribution is presented in section 5. A brief summary is given 
in section 6. 
\section{Accuracy of Future HERA Structure Function Data}
Recent measurements of the proton structure function $F_{2}(x,Q^{2})$
by the H1 and ZEUS collaborations \cite{h1f,zeusf}, based on data 
taken in 1994 with an integrated luminosity $\cal L$
 of about 3~pb$^{-1}$, 
have reached a systematic error level of about 4-5\% in the bulk 
region of the data, $10 \leq Q^{2} \leq 100$~GeV$^{2}$. Exploratory 
measurements of the very low 
$Q^{2}$ region with about 15-20\% accuracy
were presented by H1
 with 1995 shifted vertex data \cite{h1w} and by ZEUS using a rear 
calorimeter installed near the beam pipe in backward direction 
\cite{zeusp}. Based on the experience of these analyses a study has 
been made in order to estimate what might be the ultimate accuracy 
of $F_{2}$ measurements at HERA. This is a difficult task: on one hand 
one can rather easily extrapolate the present knowledge of systematic 
errors and also calculate rather 
straightforward the effect of residual 
miscalibrations on the cross section measurement. On the other hand 
there will always be local, detector dependent effects in addition 
and, furthermore, one can not simulate the results to be expected from 
innovations of the structure function analyses. For example, it is 
likely that a low electron energy calibration, much below the 
kinematic peak, can be performed reconstructing the $\pi_{0}$ mass 
or, to give another one, the region of $y$ below 0.01, which was 
considered to be not accessible due to calorimetric 
noise, may be accessed nevertheless by imposing a $p_{T}$ balance 
constraint using the electron information. Therefore this simulation 
study  may give valid estimates but the truth will be the result of 
data taking and  analysis work over many years still to come.
\newpage
For this analysis the following kinematic constraints have 
been imposed:
\begin{itemize}
 \item{$Q^{2} \geq 1$~GeV$^{2}$ which may be the limit of 
applicability of the DGLAP evolution equations at low $x$
\cite{mrs};}
 \item{ $\theta_{e} 
\leq 177^{o}$ which might be accessible with nominal energy
running even after the luminosity upgrade;}
\item{$y \leq 0.8$ a limit
arising from large radiative corrections and a small scattered 
electron energy limit $E_{e}' \geq $ few GeV due to photoproduction 
background and electron identification limitations;}
\item{$\theta_{h} \geq 8^{o}$, a hadron reconstruction limit imposed 
by the beam pipe which may differ somewhat finally.}
\end{itemize}
A number of data sets was generated as summarized in table \ref{tab1}
and illustrated in fig.1.
The maximum $Q^2$ of the data depends on the available
luminosity and might reach values of up to 
50000 GeV$^{2}$.
The generation and systematic error calculation was performed
with a numerical program written by one of us which was checked to
be in good agreement with the Monte Carlo programs used for real
data analyses. 
\begin{table}[h]
\begin{center}
\begin{tabular}{|c|c|c|c|r|r|r|}
\hline
  number & nucleon & $E_{e}$ & $E_{N}$  &
 ${\cal L}/pb^{-1}$ ¥ &  $Q^{2}_{min}$ & $Q^{2}_{max}$  \\    
\hline 
I  & proton & 27.6   & 820  & 10  & 0.5  & 100   \\  
II  & proton & 27.6   & 820  & 1000  & 100  & 50000   \\ 
III & proton & 27.6   & 400  & 200  & 100  & 20000   \\ 
IV & proton & 15.0   & 820  & 10  & 0.5  & 100   \\
V  & deuteron & 27.6   & 410  & 10   & 0.5  & 100   \\      
VI & deuteron & 27.6   & 410  & 50   & 100  & 20000   \\      
\hline
\end{tabular}
\end{center}
\caption{\label{tab1}\sf{
Summary of simulated data sets for this study, energy values are in GeV and
$Q^2$ in GeV$^2$.  
}}
\end{table}

The following systematic error sources were considered in the analysis
the effect of which is illustrated in fig.\ref{syst}:
\begin{itemize}
\item{An electron energy calibration error of 0.5\% in the backward 
region ($\theta_{e} \geq 160^{o}$) and 1\% in the central barrel and 
forward region of the detectors.}
\item{An electron polar angle uncertainty of 0.5~mrad backwards and 
1~mrad in the central part of the detector ($\theta_{e} \leq 
165^{o}$).}
\item{A 2\% uncertainty of the hadronic energy scale which is 
important at lower $y \leq 0.1$ where the kinematics cannot be 
determined solely with the electron variables $E_{e}'$ and 
$\theta_{e}$ because of divergencies of the resolution $\propto 1/y$. 
The energy scales $E_{e}'$ and $E_{h}$ may be cross calibrated by 
comparing cross section mesurements in different parts of the detector 
\cite{kb} once there is high statistics available in the barrel part,
 and using the electron and track information in the detector.}
\item{The photoproduction background may cause a 1-2\% error at large 
$y \geq 0.5$ and for $Q^{2} \leq 100~$GeV$^{2}$. This requires an 
about 10\% control of its shape and normalization which can be 
envisaged with the electron taggers, the hadronic calorimeter sections
and using tracking information in 
front of the calorimeters which suppresses the $\pi_{0}$ part of the 
contamination.}
\item{Radiative corrections can be controled to 1\%, perhaps 2\% at 
highest $y \geq 0.7$, using the hadronic and electron information 
which overconstrains the kinematics. The Monte Carlo \cite{HERAKLES}
and numerical 
calculations \cite{HECTOR} are known to be in very good agreement. 
This $F_{2}$ 
simulation assumes the radiative corrections to be performed, 
including the electroweak part which at high $y$ and $Q^{2}$ modifies 
the cross section at the $\sim 20\%$ level.} 
\item{Beam background and various efficiencies are
assumed to introduce an overall error of 2\%.}
\item{A luminosity error of 1\% is assumed.}
\end{itemize}
These systematic errors are about one half of
those presently reached in the high statistics domain of the $F_2$
measurements.
If the kinematic dependence of the correlated systematic errors is
sufficiently well known, it can be taken into account in QCD fits, 
see below.
Note in this respect that the required luminosity is not 
simply given by the statistical errors per bin but rather by the 
statistics needed for detailed systematic studies.
However there will always be residual local and higher order effects
which we represent here by a random systematic error of 1\%.
\section{Analysis Procedure} \label{ap}
The generated data were analysed using the H1 \cite{nous} and
ZEUS \cite{qcdnum} QCD fitting programs. Elsewhere in these proceedings
both programs are shown to be in good agreement \cite{compa}. In order
to simplify the analysis the data were replaced by the QCD model
(see below) so that the fits immediately converged to the minimum 
$\chi^2 = 0$ and CPU time was effectively spent only on the calculation
of the covariance matrix of the fitted parameters. The errors 
on the gluon distribution and on $\alpha_s$  are
then obtained from standard error propagation.
\subsection{QCD Model} \label{apsub1}
The QCD prediction for the $F_2$ structure function can be written as
\begin{equation} \label{analeq1}
F_2^{QCD}(x,Q^2) = F_2^{uds}(x,Q^2) + F_2^c(x,Q^2),
\end{equation}
where $F_2^{uds}$ obeys the NLO QCD evolution equations for $f = 3$ light
flavours and the charm contribution $F_2^c$ is calculated according
to \cite{riemersma}. The light flavour contribution in turn is decomposed
into a singlet and a non-singlet part:
\begin{equation} \label{analeq2}
F_2^{uds}(x,Q^2) = F_2^S(x,Q^2) + F_2^{NS}(x,Q^2).
\end{equation}
The singlet structure function is related to the singlet quark momentum
distribution, $x\Sigma = \sum_f x(q_f+\bar{q}_f)$,
which obeys an evolution equation coupled
to the gluon distribution $xg$. The main contribution to $F_2^{NS}$
comes from the difference of up and down quarks and antiquarks:
$x\Delta_{ud} = x( u + \bar{u}) - x(d + \bar{d})$. We remark here that
$\Delta_{ud}$ is constrained by the difference $F_2^p - F_2^d$ of proton
and deuteron structure functions.

At the input scale $Q^2_0 = 4$ GeV$^2$ the parton distributions were
parametrised as
\begin{eqnarray} \label{analeq3}
xG(x,Q_0^2) & = & A_G x^{B_G}(1-x)^{C_G} \nonumber \\
x\Sigma(x,Q_0^2) & = &  A_S x^{B_S}(1-x)^{C_S}(1+D_{S}x+E_S \sqrt{x}) \\
x\Delta_{ud}(x,Q_0^2) & = & A_{NS} x^{B_{NS}}(1-x)^{C_{NS}}. \nonumber
\end{eqnarray}
The input parameters for the gluon and the singlet distributions were
obtained from a fit to the simulated 
data whereas the non-singlet parameters and their
uncertainties were taken from \cite{ZEUSQCD}.
The input value of $\alpha_{s}$ was set to
$\alpha_{s}(M_{Z}^2)$ = 0.113
corresponding to 
$\Lambda_{_{\overline{\rm MS}}}^{(4) } = 263$ MeV \cite{MARCALAIN}.

\subsection{Definition of the $\chi^2$ and Fit Procedure} \label{apsub2}
The two
fitting programs of H1 and ZEUS have been used in parallel and all
important numbers were cross checked. The
ZEUS program uses a step by step ( \`a la Runge Kutta)
 procedure to solve the DGLAP evolution equations. The
H1 program projects the  DGLAP equations on
 a functional basis where they are solved
exactly \cite{nous,nous2}.
Both programs use MINUIT to make the fitting. In addition the H1 program
has the possibility to use an independent 
  set of routines (called LSQFIT) which performs
a least chisquare fit. In LSQFIT the $\chi^{2}$
function to be minimised is
recognized to be the sum of the square of deviations 
and the derivatives of the deviations are computed by finite differences.
Both MINUIT and LSQFIT can compute the second order derivatives of the
 $\chi^2$ with respect to the parameters: these may be used for the
error computation as  will be shown in the following.

The $\chi^2$ is defined as
\begin{equation} \label{chi2}
\chi^2 = \sum_{i}
 \biggl(\frac{F_i(p,s)-f_i}{ \Delta f_i}\biggr)^2 + \sum_{l} (s_l)^2
\end{equation}
where $F_i$ is the model prediction, $f_i$ the measured $F_2$ value,
$\Delta f_i$ its statistical error and the sum runs over
all data points ($i$). In addition to the set of 
parton distribution parameters
$\{p\}$, including $\alpha_s$, we have introduced the parameter set
 $\{s\}$
which takes into account the systematic errors of the measurements.
The relation between the model prediction and the QCD prediction for
$F_2$ is written as:
\begin{equation} \label{sysdef}
F_i(p,s) = F_i^{QCD}(p)~~ ( 1 - \sum_l  s_l \Delta_{li}^{syst})
\end{equation}
where $\Delta_{li}^{syst}$ is the relative 
systematic error on data point
($i$) belonging to the source ($l$).
 We assume  that the parameters $s_l$
are gaussian distributed with zero mean and unit variance so that
the $\Delta_l$ correspond to
a one standard deviation systematic error\footnote{Asymmetric errors
 can be taken
into account by adding terms quadratic in $s_l$ in eq.~\ref{sysdef}.}.

\subsection{Systematic Error Evaluation} \label{apsub3}

Given the $\chi^2$ definition of the previous section there are essentially
three methods to evaluate the systematic errors on the fitted parameters:
\begin{itemize}
\item {Repeat the fit with several values of the systematics variables
      $s_l$, either chosen at random or
      giving to each variable in turn the value 1. The systematic errors
      are then obtained by adding all the deviations
      from the central value in quadrature.}
\item {Leave the systematic parameters fixed to zero but propagate
      the errors on $s_l$ (assumed to be 1) to the covariance 
      matrix of the fitted parameters \cite{nous3}.
If the deviations are linear functions of the systematic variables
it is easy to compute directly the errors from the second derivatives
of the $\chi^2$.
Let us introduce the following matrices:
\begin{equation}
  M=\sum_{i} {\partial {F_i} \over \partial p }
         {\partial {F_i} \over \partial p } {1 \over {\Delta m_i^2}}
  \approx 1/2{\partial^2  \chi^2 \over {\partial p \partial p}} 
\end{equation}
\begin{equation}
  C=\sum_{i} {\partial {F_i} \over \partial p }
         {\partial {F_i} \over \partial s } {1 \over {\Delta m_i^2}}
  \approx 1/2{\partial^2  \chi^2 \over {\partial p \partial s}} 
\end{equation}
  $ V^{stat} = M^{-1}$ is the $\{p\}$ statistical error matrix 
and $C$ is the matrix which
expresses statistical and systematic correlations.
One can show that \cite{nous3}
\begin{equation} \label{sysvar}
 V^{syst} =  M^{-1} C C^T M^{-1}
\end{equation}
is the $\{p\}$ systematic error matrix. The 
LSQFIT program determines these matrices 
and also the function error bands. 
For MINUIT  a
 fit has to be performed where  the systematic parameters
are left free  and the inverse
of the resulting covariance matrix contains
 the matrices $M$ and $C$. Reinverting
M and using eq.(\ref{sysvar})  yields the statistical
 and systematic errors
in case all systematic parameters are kept fixed.}
\item { The $\chi^2 =\chi^2_{min}+1$ method\\
If the correlations between parameters are big and/or the dependence
of the deviations with respect to the parameters is highly 
non--linear,
it is more appropriate to compute the error on a specific
 parameter by considering an
  increase of the $\chi^2$ by one, all the other
parameters being optimised.
Both MINUIT and LSQFIT can provide
 this calculation. This method has been used also to draw error
 bands with LSQFIT which is faster than the MINOS option of MINUIT.
In this case the value of the function itself at some
fixed x and $Q^2$ point is taken as a parameter. Then the equation say
$G(x,Q^2)= G$ is used to eliminate the more sensitive parameter.}
\end{itemize}

The three  different methods and the two different programs have
been compared in detail leading to consistent results. 
Fits were performed on data  randomly 
offset both statistically
$and$ systematically. Taking  as the error
the r.m.s.\ of the fitted  
 $\alpha_s$ values
 this appeared to be in agreement with the standard
 error calculation.
This method gives the most reliable error 
estimate, but it is
clearly too elaborate to be of practical use in a study like the
one undertaken here. 

\subsection {Fitting the Systematics}
If the kinematical dependence of a
systematic error, like the $1/y$ behaviour of
the electron energy scale uncertainty, is well known, a
 contribution, 
 $ \sum_{l} s_{l}^2 $, can be added
to the $\chi^2$ and  a fit can be performed
determining an extended set of  parameters
$\{p,s\}$.
The interest of such a procedure is obviously that here 
 full knowledge of the experiment enters 
to improve the measurement accuracy. Such a procedure was adopted
in \cite{MARCALAIN} to reduce the influence of the main 
experimental errors, the magnetic field calibration of the BCDMS 
spectrometer for example, on the value and error of $\alpha_{s}$.
The method to find the resulting errors is practically
the same as for the statistical error treatment.
LSQFIT and MINUIT deliver the complete error matrix which
in the case of MINUIT  is exactly the one used in
method 2.
\section{Results on $\alpha_s$} \label{secas}
\subsection{Introduction}
The data and the fitting procedures as described above were used to
determine the expected error of $\alpha_{s}(M_{Z}^{2})$, and of $xg$ 
in the subsequent section. Three types of fits were performed:
\begin{itemize}
\item {A - Fits to HERA proton data alone.}
\item {B - Fits to HERA proton and deuteron data.}
\item {C - Fits to HERA proton data with inclusion of fixed target
      data, in most of the cases those
       from SLAC \cite{SLAC} and BCDMS \cite{BCDMS}.}
\end{itemize}
In the  fits the systematic error parameters were left free.
 The input values
$s_l = 0$ were always reproduced while the input
 errors $\Delta_{s_l} = 1$
were typically reduced by a factor of two. The correlation coefficients
between the systematic error parameters were well below unity.
In the following we denote the error on $\alpha_s$ from these fits
by $\Delta \alpha_{fit}.$
The statistical error
($\Delta \alpha_{stat}$) and the systematic 
error ($\Delta \alpha_{syst}$) for fixed systematic errors 
were calculated from the covariance matrix as described  above.
\subsection{$\alpha_s$ with HERA Data only}
As a starting point of the investigation fits were made
to HERA high energy proton data alone
 (sets I and II in table~\ref{tab1}). 
The low $Q^2$ sample (set I) covers an $x$ range of
$1.4 \times 10^{-5} < x
< 4.3 \times 10^{-2}$ whereas the high $Q^2$
sample (set II) covers $2.4 \times 10^{-3} < x < 0.65.$ 
For the nominal fits integrated luminosities
of 10 and 500 pb$^{-1}$
were assumed  for the low and the high $Q^2$ data set respectively.

The strong coupling constant  and
the parameters describing the input singlet (5 parameters) and gluon
distributions (3 parameters) at $Q^2_0 = 4$ GeV$^2$ 
were left free in the fit. The gluon 
normalization was calculated
by imposing the momentum sumrule. The 
non-singlet contribution to $F_2$ 
was kept fixed since it is not well constrained by proton data alone.

Besides $\alpha_s$ and the parton distribution parameters
five systematic error parameters were introduced as described
in section~3. In addition the assumed random systematic error
was added in quadrature to the statistical error.

Recent analyses of ZEUS \cite{ZEUSQCD} and H1 \cite{h1f}
$F_2$ data have shown that perturbative QCD might be applicable 
down to $Q^2 \simeq 1$ GeV$^2$ at very low $x$, see also \cite{mrs}.
In figure~\ref{mbplot1}a and table~\ref{astab} 
 the  $\alpha_s$ error is given
as a function of  $Q^2_{c}$ which is the lowest $Q^2$ considered in
the fit. The statistical error
(which includes the 1\% random systematic error) increases from
$\Delta \alpha_{stat} = 0.0024$ to 0.0053 with increasing $Q^2_{c}$.
When all systematic errors are fitted, $\Delta \alpha_{fit}$ is
almost identical to the statistical error for $Q^2_{c} = 1$
GeV$^2$ but increases more rapidly to 0.0075 at $Q^2_{c} = 8$
GeV$^2.$ When the systematics are not fitted, their contribution
to $\Delta \alpha_s$ rises very strongly to about 0.012 above
$Q^2_{c} = 2$ GeV$^2.$
The same tendency is observed when all the systematic errors are
scaled down by a factor of two (figure~\ref{mbplot1}b) though,
as expected, the  uncertainty of $\alpha_s$ is reduced
almost by a factor of two.

To investigate the impact of the high $Q^2$ data on the result,
the luminosity of dataset II 
was varied between ${\cal L} =$ 10 and 1000 pb$^{-1}$. 
For a $Q^2_{c}$ cut of 1~GeV$^2$ 
the variation of the $\alpha_s$ errors
with $L$ is fairly modest as shown in figure~\ref{mbplot2}a 
and table~\ref{astab}. 
However, the dependence on the high $Q^2$ luminosity
becomes stronger if  the $Q^2_{c}$ cut is raised. This is
illustrated in 
figure~\ref{mbplot2}b for $Q^2_{c} = 3$ GeV$^2.$ Here a factor of 10
increase in luminosity decreases the $\alpha_s$  errors by about 40\%.
On the other hand, increasing the luminosity
of the low $Q^2$ sample (dataset I)
from 10 to 50 pb$^{-1}$ gave an insignificant
improvement ($\approx 10\%$) on the uncertainty in $\alpha_s$.

An improvement of the result is obtained if the lower energy data
are included, sets III and IV in table 1. For example, for 
$Q^{2}_{c} = 3$~GeV$^{2}$ 
the nominal data set yields an error of 0.0061, see table \ref{astab}
 while 
the inclusion of the lower energy data reduces that error to 0.0046, 
if in both cases the systematics is fitted.

As described above, the nonsinglet distribution is input 
to the fit of the proton data. 
Taking its uncertainty from the QCD analysis
of~\cite{ZEUSQCD} a contribution of about 0.004 is estimated to
the uncertainty on $\alpha_s$. However, when both proton and deuteron
data are available the nonsinglet contribution is constrained by the
difference $F_2^p - F_2^d$ and the 0.004 error gets eliminated.
Therefore, a low and a high $Q^2$ deuteron data sample (set V and VI
in table~\ref{tab1}) with modest lumionosities were
included in the fit.  Apart from the highest $Q^2$ region
these data have roughly
the same kinematic coverage as the proton data.

In the combined proton and deuteron QCD fit
the three parameters which describe the nonsinglet input distribution
were left free. Furthermore one normalization parameter and five 
independent
systematic parameters for the deuteron data were added.
The resulting  $\alpha_s$ errors are given in table~\ref{astab}
for a $Q^2_{c}$ cut of 3 GeV$^2.$ It is seen that the 
error on $\alpha_s$
is reduced by about 25\% compared to the corresponding
fit on proton data only although the number of fit parameters had
to be increased and the non-singlet distribution is as well fitted. 

\subsection{Inclusion of High $x$ Fixed Target Experiment Data}
The published QCD analysis \cite{MARCALAIN} of SLAC and
BCDMS  proton and deuteron structure function
 data yielded an
experimental error of 0.003 on $\alpha_s(M_Z^2)$. The natural 
question to be answered is whether
the combination of the low $x$ and high $Q^{2}$ HERA data with
 the fixed target experiment data can improve this result
significantly.
 
As a first necessary step it was studied if
our fits can reproduce the error quoted above. 
The following conditions were applied to mimic the analysis of
\cite{MARCALAIN} as closely as possible:
\begin{itemize}
\item {A cut of $W^2 > 10$ GeV$^2$ was imposed to effectively
      remove the region at high $x$ and low
      $Q^2$ dominated by higher twist effects.}
\item {The parameters of the input singlet, gluon and
      nonsinglet distributions were left free except
      $B_S$ and $B_G$ which describe
      the low $x$ behaviour of $x\Sigma$ and $xG$
      (the SLAC/BCDMS data extend only down to $x = 0.07$).}
\item {One normalisation parameter was kept fixed (BCDMS
      deuteron) whereas those of the remaining three datasets
      were left free. In addition two systematic parameters for the
      BCDMS data were left free.}
\item {No momentum sumrule was imposed.}
\item {Being interested in the derived error only, the SLAC and BCDMS 
      data were replaced by the model input.}
\item {The quoted errors of ref~\cite{MARCALAIN} correspond to an
      increase of the $\chi^2$ by 9 units which was taken into 
      account in the estimate of the statistical errors.} 
\end{itemize}
The fit defined above on the SLAC/BCDMS data alone yielded as a result
$\Delta \alpha_s = 0.0030,$ exactly as published.

Using all high energy HERA proton data in addition
the error on 
$\alpha_s$ was reduced to $\Delta \alpha_{fit} = 0.0016$
with $Q^2_{c} = 3$~GeV$^2$. Adding the latest data of the
NMC experiment with a preliminary treatment of the systematic
errors of this measurement reduces this number to 0.0013. 
In table~\ref{astab} results are given on $\Delta \alpha_s$ for
various choices of the $Q^2_{c}$ cut, the size of the systematic
errors and the luminosity of the high $Q^2$ data sample. It turns
out that $\Delta \alpha_{fit}$ ranges from about 0.001 to 0.002
and is thus fairly insensitive to these choices. 
Compared to fits on HERA proton data alone, the error on $\alpha_s$
is much less sensitive as to whether the systematic parameters are
left free or kept fixed and the dependence on the
minimum $Q^{2}$ is less severe.
With fixed systematic errors the combined statistical
and systematic error ranges from $\Delta \alpha_s = 0.002$ to about
0.003.
\begin{table}
\center
\begin{verbatim}
  #===#============#==========#==============================#
  |Fit| L(LQ) L(HQ)|sfac qmin |   sigf   stat   syst  total  |
  #===#============#==========#==============================#
  | A |  10   500  | 1.0  1.0 |  .0024  .0021  .0052  .0056  |
  | A |  10   500  | 1.0  2.0 |  .0050  .0036  .0107  .0113  |
  | A |  10   500  | 1.0  3.0 |  .0061  .0041  .0114  .0120  |
  | A |  10   500  | 1.0  5.0 |  .0069  .0047  .0114  .0123  |
  | A |  10   500  | 1.0  8.0 |  .0075  .0052  .0105  .0117  |
  |   |            |          |                              |
  | A |  10   500  | 0.5  1.0 |  .0015  .0013  .0024  .0028  |
  | A |  10   500  | 0.5  2.0 |  .0033  .0026  .0055  .0061  |
  | A |  10   500  | 0.5  3.0 |  .0040  .0031  .0060  .0067  |
  | A |  10   500  | 0.5  5.0 |  .0046  .0035  .0060  .0070  |
  | A |  10   500  | 0.5  8.0 |  .0050  .0040  .0054  .0067  |
  +---+------------+----------+------------------------------+
  | A |  10    10  | 1.0  1.0 |  .0028  .0023  .0045  .0050  |
  | A |  10   100  | 1.0  1.0 |  .0026  .0022  .0048  .0053  |
  | A |  10  1000  | 1.0  1.0 |  .0023  .0020  .0053  .0057  |
  |   |            |          |                              |
  | A |  10    10  | 1.0  3.0 |  .0096  .0080  .0126  .0149  |
  | A |  10   100  | 1.0  3.0 |  .0073  .0055  .0123  .0134  |
  | A |  10  1000  | 1.0  3.0 |  .0057  .0037  .0110  .0116  |
  #===#============#==========#==============================#
  | B |  10   500  | 1.0  3.0 |  .0043  .0030  .0087  .0092  |
  #===#============#==========#==============================#
  | C |  10   500  | 1.0  1.0 |  .0014  .0011  .0031  .0033  |
  | C |  10   500  | 1.0  8.0 |  .0016  .0012  .0032  .0034  |
  |   |            |          |                              |
  | C |  10   500  | 0.5  1.0 |  .0011  .0009  .0018  .0020  |
  | C |  10   500  | 0.5  8.0 |  .0014  .0011  .0018  .0022  |
  +---+------------+----------+------------------------------+
  | C |  10    10  | 1.0  3.0 |  .0020  .0018  .0022  .0028  |
  | C |  10   100  | 1.0  3.0 |  .0017  .0015  .0023  .0027  |
  | C |  10  1000  | 1.0  3.0 |  .0015  .0011  .0030  .0032  |
  #===#============#==========#==============================#
\end{verbatim}
\caption{\sf{Errors on $\alpha_s$} for fits with HERA proton data only
 (A), proton and deuteron data (B) and combinations of simulated HERA
 proton data with fixed target experiment data (C)
 from SLAC and BCDMS, see text.}
\label{astab}
\end{table}
\subsection{Double Logarithmic Scaling and the Error of $\alpha_{s}$}
With high precision data the low $x$ behaviour of 
$F_{2}$ will be much better understood. If the data further support
the double logarithmic approximation \cite{bf} 
of the  low $x$, large $Q^{2}$ behaviour of 
$F_{2}$, then a precision of $\alpha_{s}$ to 0.001 or even better can be
reached with HERA data alone. Already with the present H1 
data only, such an analysis \cite{atm} lead to a value of
 $\alpha_{s}(M_{Z}^{2}) = 0.113 \pm 0.002 (stat) \pm 0.006 (syst)$.
The advantage 
of this approach is obvious as it likely requires only the low $x$ 
data of HERA and depends on two scale parameters, $Q^{2}_{o}$, 
$x_{o}$, a normalization constant and $\alpha_{s}$ only. This is in 
contrast to the QCD analysis of HERA and fixed target data,
considered here, which has to
include the full parametrization of two nonsinglet, the singlet and 
the gluon distribution leading to typically 15 parameters to be 
simultaneously controled. Further theoretical understanding of the
double scaling approach is necessary, however.
\section{Determination of the Gluon Distribution}
\subsection{HERA Proton Data Only}
The previous determinations of the gluon distribution from the
scaling violation of $F_{2}$ at low $x$ by H1
\cite{h1f,h1g} and ZEUS \cite{zeusf,ZEUSQCD} were performed 
combining the HERA results with fixed target data and treating
$\alpha_{s}$ as an extra parameter. Since the scaling violations 
essentially are proportional to the product of $\alpha_{s} \cdot xg$,
a large data range in $x$ and $Q^{2}$ is required to disentangle these 
two basic quantities.

 Fig.\ref{gluon1} shows the result of a QCD fit to 
the simulated $F_{2}$ data, sets I and II in table 1,
 without fitting the systematics but
determining  $xg$ and the singlet 
distribution $x \Sigma$. In this fit $\alpha_{s}$ was fixed considering an  
$\alpha_{s}$ 
uncertainty of 0.005  for the calculated
 gluon error. The inner dark error band is the statistical
error while the total error is shown as the outer grey band for all
$Q^{2}$ values. The error bands were drawn using the LSQFIT routines 
and the  $\chi^2 =\chi^2_{min}+1$ method.
 At $Q^{2} = 20$~GeV$^{2}$ and $x=0.0001$ the total 
error amounts to 11\% somewhat better than the present result which
included the fixed target experiments.  If  $\alpha_{s}$ is allowed 
to vary and the systematics  is fitted 
as described in section 3 the gluon determination gets very accurate 
with an estimated error of 3\% at the same $x$ and $Q^{2}$ values. This
is illustrated in  figure~\ref{gluon2}.
\subsection{HERA Proton and Deuteron Data}
The consideration of deuteron data of ${\cal L} = 50$~pb$^{-1}$
allowed finally to perform a complete fit of all distributions. In 
particular one has to notice that the non--singlet distributions are
not well constrained with proton data only. In section 3 the up--down 
quark distribution difference was introduced as the
 non--singlet quantity to 
be determined assuming the strange distribution was fixed, see 
\cite{ZEUSQCD}. Generally there are two non--singlet distributions 
which may be written as $u^{+}=u+\bar u - \Sigma/3$ and similarly $d^{+}$.
The deuteron data allow to determine these distributions. Fig.\ref{nons}
 shows the calculated accuracy of the $u^{+}$ distribution 
at $Q_{o}^{2} = 2$~GeV$^{2}$ from a complete fit to all distributions, 
$\alpha_{s}$ and with free systematic error parameters. This 
represents an interesting result as apparently the non--singlet 
distributions will be measurable  accurately
down to very low $x$, and the 
predicted weak $Q^{2}$ dependence can be verified. The gluon 
belonging to this fit is determined with an error of only 1-2\% at the
$x,Q^{2}$ point chosen for comparison.  

Further studies of the 
importance of HERA deuteron data are necessary. For example,
this data will have 
an  additional  few \% uncertainty due to shadowing 
corrections at low $x$ which was neglected here. 
Constraints on the non--singlet distributions are  available 
already from the fixed target deuteron data, but at higher $x$.
A more thorough discussion of these aspects has been beyond the scope 
of this study. Independently of theoretical preassumptions, however, 
one may regard this data as important to measure the low $x$
behaviour of the up-down quark distribution difference which requires 
modest luminosity only. With larger luminosity ${\cal L} \simeq 
50$~pb$^{-1}$ it will help to 
decompose the flavour contents of the nucleon as was discussed 
already ten years ago \cite{jtmr}.
\section{Summary}
According to this analysis HERA will have an important impact on the 
measurement of $\alpha_{s}$ and the gluon distribution. A precision 
near  0.001 for $\alpha_{s}(M_{Z}^{2})$ and of almost 1\% for $xg$
is in reach if
\begin{itemize}
\item{$F_{2}$ measurements in the full HERA range
will become available with systematic and 
statistical errors of a few \% only,}
\item{the systematic errors are thoroughly studied at the per cent 
level as functions of $x$ and $Q^{2}$ such that their gross effects 
can be absorbed in the QCD analysis,}
\item{and the HERA data can be combined reliably with the fixed
target experiment results on $F_{2}$.}
\end{itemize}
Such an accuracy represents a great challenge for the experimental
programme at HERA. The HERA data should be complete, i.e. comprise a 
high luminosity data set with ${\cal L} \geq 300$~pb$^{-1}$ and modest 
luminosity data sets: i) at lowered proton energy 
${\cal L} \geq 50$~pb$^{-1}$ 
to reach highest $x$, as 
close as possible to the fixed target data region,  ii) at
lowered electron 
beam energy ${\cal L} \geq 3$~pb$^{-1}$
to cover the $x$ dependence at smallest $Q^2$ and  iii) possibly also
 deuteron data with ${\cal L} \simeq 50$~pb$^{-1}$. Note that no 
 attempt was made to optimize these luminosity values, in particular, 
 since the necessary level of systematic error control is 
 competing with the 
 requirements coming from simple statistical error considerations.

Completion of this programme also requires a major theoretical 
effort to calculate the 3-loop coefficient 
functions since the present theoretical uncertainty of $\alpha_{s}$
of $\sim 0.005$ \cite{jret}
exceeds most of the estimated experimental  $\alpha_{s}$ errors
discussed in this study. It is obvious that a precision at the
level of 0.002 
 for $\alpha_{s}(M_{Z}^{2})$ will lead to a very precise 
study of its $Q^{2}$ dependence and resolve the question of the
compatibility of the deep inelastic $\alpha_{s}$
 values with those from $e^{+}e^{-}$ scattering. 
\\
\\
{\bf Acknowledgement} We would like to thank J.~Bl\"umlein, 
Th.~Naumann, S.~Riemersma, A.~Vogt and F.~Zomer for interesting 
discussions on various aspects of this study.

\newpage
\begin{figure}[hb]\centering
 \begin{picture}(160,500)
\put(-150,-80){
\epsfig{file=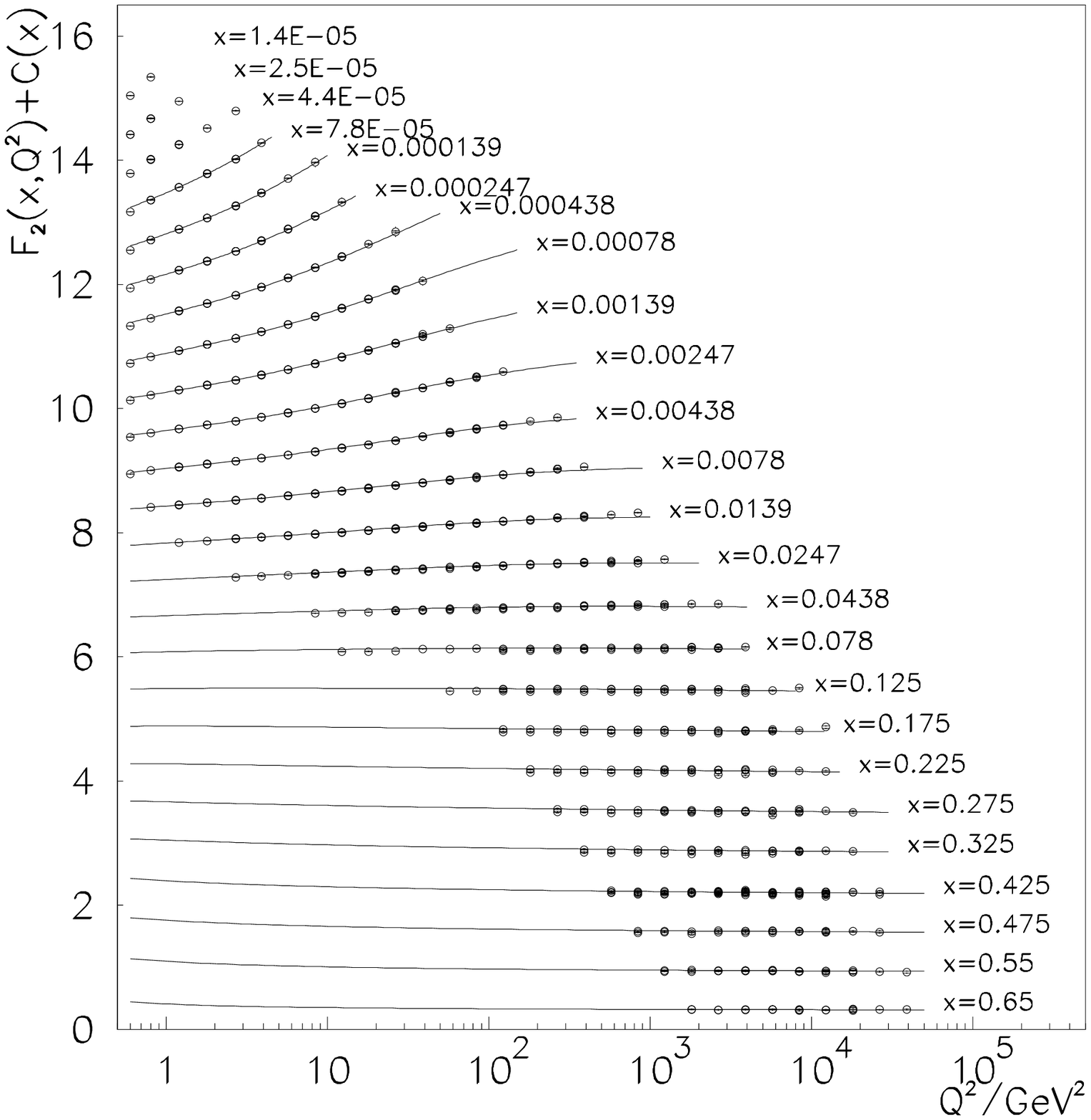,bbllx=0pt,bblly=0pt
,bburx=557pt,bbury=792pt,width=14cm,height=21cm}}
\end{picture}
        \caption{
        \sf{Simulated structure function data sets. The huge 
        luminosity of 1~fb$^{-1}$ will lead to precise data even at
        very high $Q^{2}$. For $Q^{2} \geq 10000$~GeV$^{2}$ about 
        2000 events may be available. The largest bins shown are made 
        with 20-50 events. With ${\cal L} = 10$~pb$^{-1}$ for 
        $Q^{2} \geq 0.5$~GeV$^{2}$ about $10^{7}$ events are occuring
        which will be prescaled at lowest $Q^{2}$.  The curve 
        represents a NLO QCD fit. The high $x$, low $Q^{2}$ region 
        can not be accessed with HERA but
        is almost completely covered by the fixed target experiment
        data, not shown here.}}
        \protect\label{prop}
 \end{figure} 
\newpage
\begin{figure}[h]\centering
\begin{picture}(160,500)
\put(-150,-100){
\epsfig{file=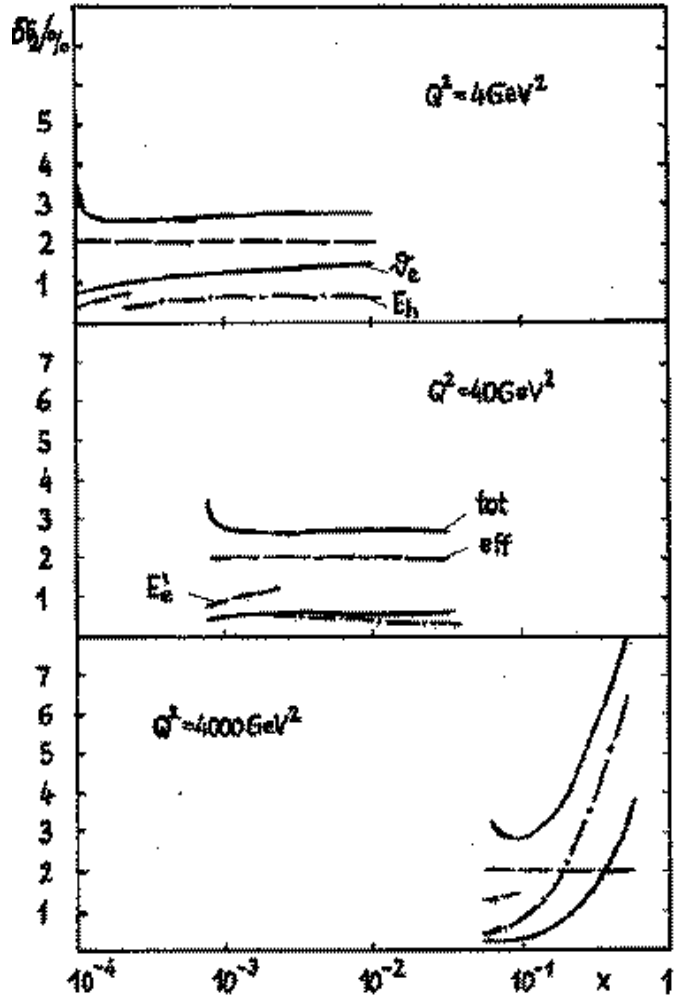,bbllx=0pt,bblly=0pt
,bburx=557pt,bbury=792pt,width=14cm,height=21cm}}
\end{picture}
        \caption{
        \sf{
        Estimated systematic errors of the $F_{2}$ 
        measurement for different $Q^{2}$ as a function of $x$.
        Dashed line: effect of error on the scattered electron energy
    $E_{e}'$, dashed-dotted line: effect of error on the hadronic 
    energy scale $E_{h}$, solid line: effect of error on the polar 
    angle $\theta_{e}$; long dashed line: 2\% efficiency error. Not 
    drawn are the effect of photoproduction background at high $y$
    and the radiative correction error. Both have been added to the 
    other error sources which gives a total error drawn as the upper 
    solid line.}}
        \protect\label{syst}
 \end{figure}
\begin{figure}[ht] 
\centerline{
\psfig{figure=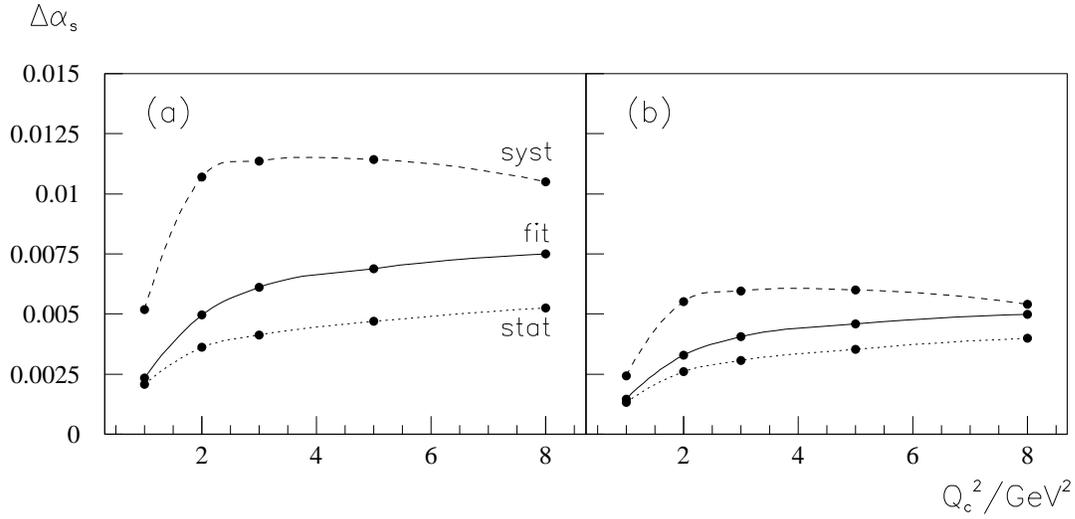,width=16cm}}
\caption{\sf{The error on $\alpha_s(M_Z^2)$ from 
fits to the HERA high energy
proton data as a function of $Q^2_{c}$: (a) with full systematics 
included; (b) with systematics further
reduced by a factor of two. The (dotted,
solid, dashed) curves correspond to the errors ($\Delta \alpha_{stat}$,
$\Delta \alpha_{fit}$, $\Delta \alpha_{syst}$) described in the text.}}
\protect\label{mbplot1}
\end{figure}
\begin{figure}[hb] 
\centerline{
\psfig{figure=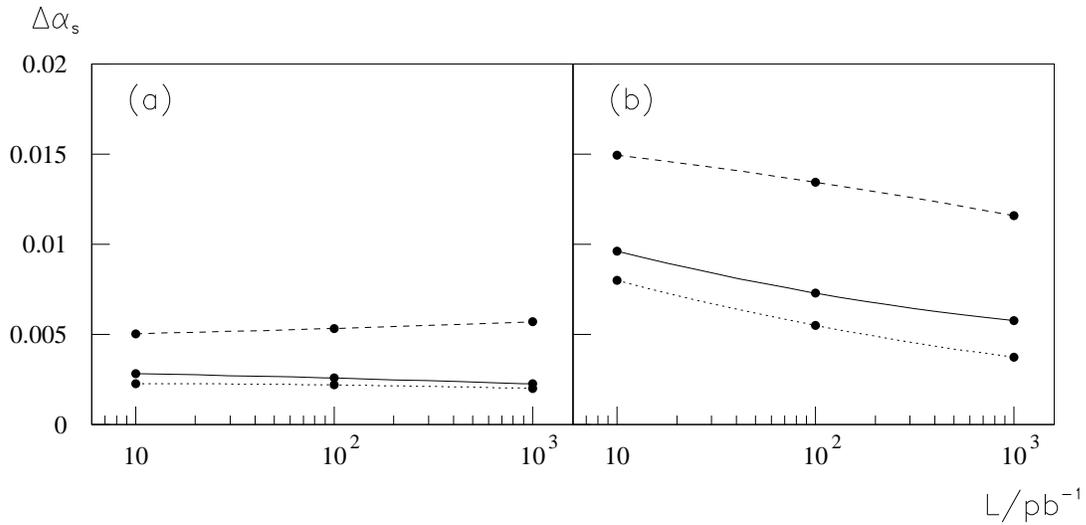,width=16cm}}
\caption{\sf{The error on $\alpha_s(M_Z^2)$ as a  function of the luminosity
of the high $Q^2$ sample: 
(a) for $Q^2_{c} = 1$ GeV$^2$; (b) for $Q^2_{c} = 3$ GeV$^2$.
The (dotted,
solid, dashed) curves correspond to the errors ($\Delta \alpha_{stat},$
$\Delta \alpha_{fit},$ $\Delta \alpha_{syst}$) described in the text.}}
\protect\label{mbplot2}
\end{figure} 
\begin{figure}[h]\centering
 \begin{picture}(160,300)
\put(-150,-90){
\epsfig{file=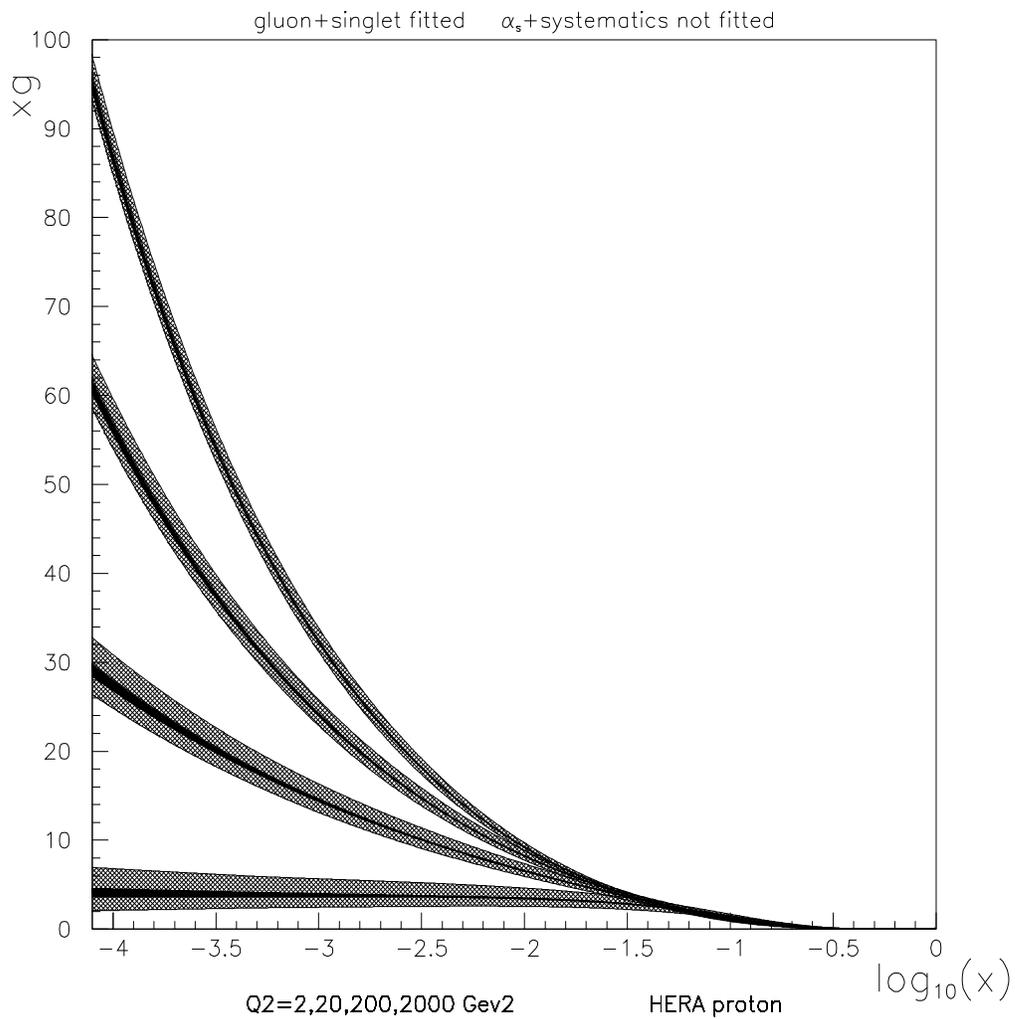,bbllx=0pt,bblly=0pt
,bburx=557pt,bbury=792pt,width=14cm,height=21cm}}
\end{picture}
        \caption{
        \sf{Determination of the gluon distribution using future $F_{2}$
        data from electron-proton scattering with fixed  
         systematic error parameters. The inner band is the statistical
        error. 
         Note that for simplicity the gluon is 
        shown  outside the allowed region of
        $x \leq Q^{2}/10^{5}$.}}
        \protect\label{gluon1}
 \end{figure}
\begin{figure}[h]\centering
 \begin{picture}(160,300)
\put(-150,-90){
\epsfig{file=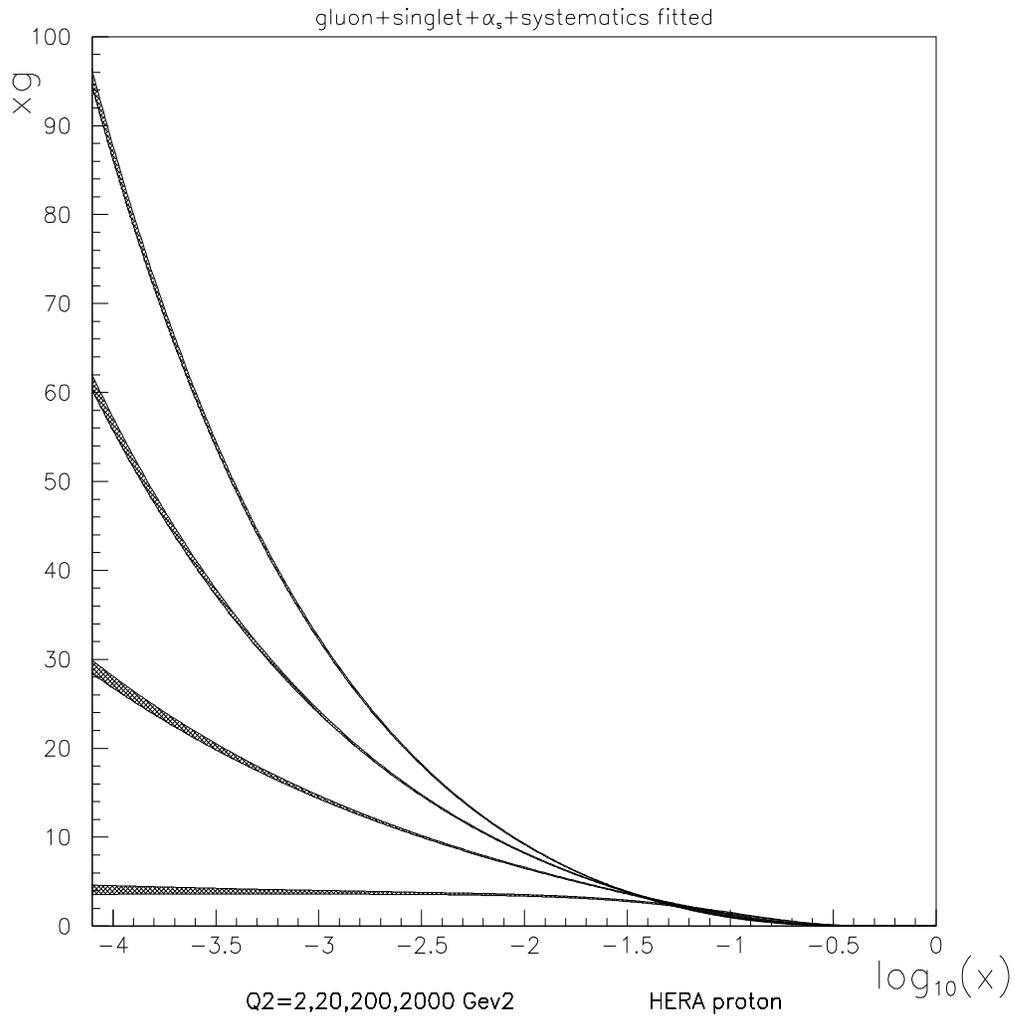,bbllx=0pt,bblly=0pt
,bburx=557pt,bbury=792pt,width=14cm,height=21cm}}
\end{picture}
        \caption{
        \sf{Determination of gluon distribution using future $F_{2}$
        data from electron-proton scattering with 
        fitted systematic error parameters. 
         Note that for simplicity the gluon is 
        shown  also  outside the allowed region of 
 $x \leq Q^{2}/10^{5}$.}}
        \protect\label{gluon2}
 \end{figure}
\begin{figure}[b]\centering
 \begin{picture}(160,300)
\put(-150,-90){
\epsfig{file=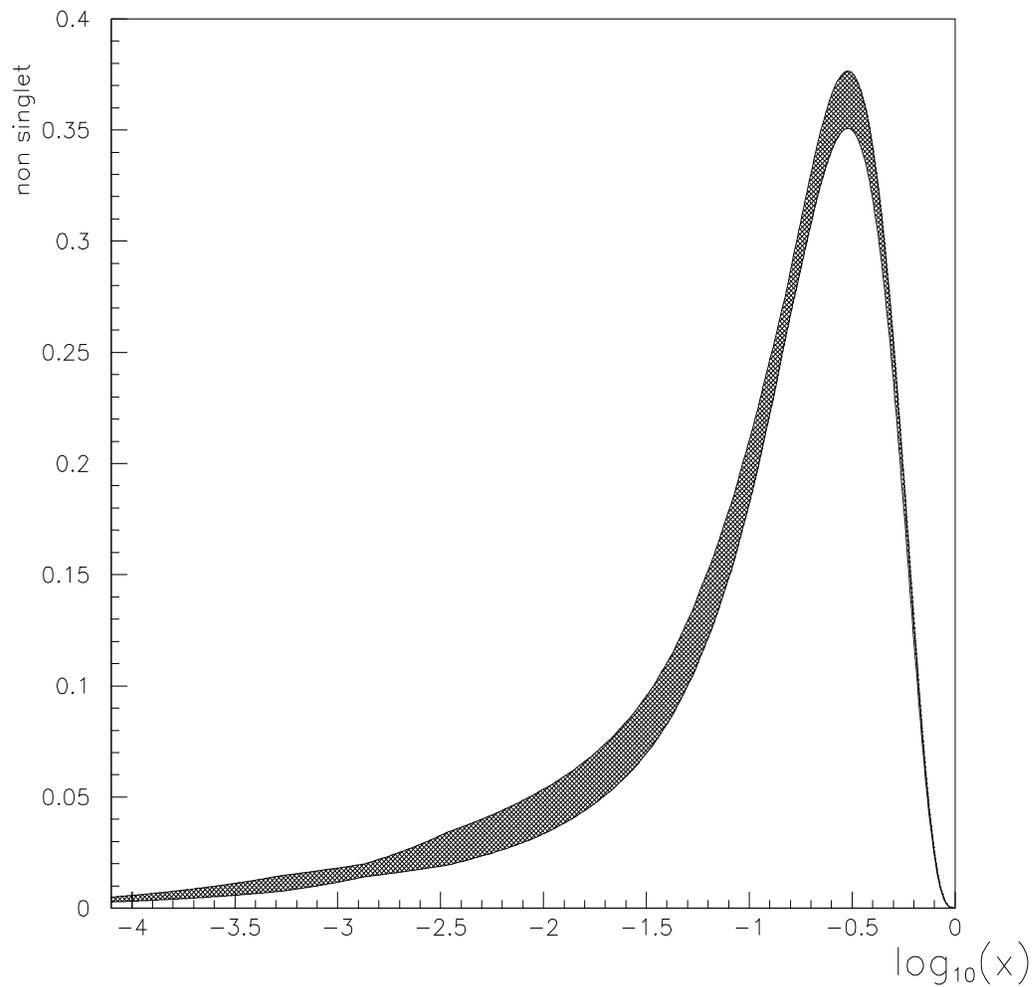,bbllx=0pt,bblly=0pt
,bburx=557pt,bbury=792pt,width=14cm,height=21cm}}
\end{picture}
        \caption{
        \sf{Determination of the non-singlet distribution $u^+ =
         u+\bar u - \Sigma/3$ from a QCD fit to the simulated proton and
         deuteron data (sets I,II and V,VI in table 1).}}
        \protect\label{nons}
 \end{figure}¥ 
\end{document}